\documentclass{ifacconf}
\usepackage{xcolor}
\usepackage{color}
\usepackage{graphicx}
\usepackage[noend]{algpseudocode}
\usepackage{CJK}    
\usepackage{algpseudocode}
\usepackage{algorithmicx,algorithm}
\usepackage{subfigure}
\usepackage{amsfonts}
\usepackage{algorithmicx}
\usepackage{mathtools} 
\usepackage{amsmath}
\usepackage{amssymb}
\usepackage{booktabs}
\usepackage{float}
\usepackage{verbatim}
\usepackage{algorithm}  
\usepackage{algpseudocode}  
\usepackage{amsmath}  
\usepackage{graphicx}      
\usepackage{natbib}        
\usepackage{stfloats}

\usepackage{amsmath}
\DeclarePairedDelimiter{\abs}{\lvert}{\rvert}
\newcommand{\norm}[1]{\lVert #1 \rVert}

\definecolor{dogwoodrose}{rgb}{0.84, 0.09, 0.41}

\definecolor{azure}{rgb}{0.0,0.5,0.9}

\definecolor{ygo}{rgb}{1, 0.25, 0.05}
\definecolor{blue}{rgb}{0.00002 0 1}
\definecolor{persimmon}{rgb}{0.93, 0.35, 0.0}
\definecolor{mangotango}{rgb}{1.0, 0.51, 0.26}

\usepackage[normalem]{ulem}

\begin{document}
\begin{frontmatter}

\title{Shepherding Control for Separating a Single Agent from a Swarm\thanksref{footnoteinfo}} 

\thanks[footnoteinfo]{This work was supported by JSPS KAKENHI Grant Number JP21H01352.}

\author[OU]{Yaosheng Deng}
\author[OU]{Masaki Ogura}
\author[OU]{Aiyi Li}
\author[OU]{Naoki Wakamiya}
\address[OU]{Graduate School of Information Science and Technology, Osaka University, Suita, Osaka 565-0871 Japan (e-mail: ys-deng@ ist.osaka-u.ac.jp).}

\begin{abstract}                
In this paper, we consider the swarm-control problem of spatially separating a specified target agent within the swarm from all the other agents, while maintaining the connectivity among the other agents. We specifically aim to achieve the separation by designing the movement algorithm of an external agent, called a shepherd, which exerts repulsive forces on the agents in the swarm. This problem has potential applications in the context of the manipulation of the swarm of micro- and nano-particles. We first formulate the separation problem, where the swarm agents (called sheep) are modeled by the Boid model. We then analytically study the special case of two-sheep swarms. By leveraging the analysis, we then propose a potential function-based movement algorithm of the shepherd to achieve separation while maintaining the connectivity within the remaining swarm. We demonstrate the effectiveness of the proposed algorithm with numerical simulations. 
\end{abstract}


\end{frontmatter}

\section{Introduction}

Swarms are formed by various organisms, from simple bacteria to more advanced mammals. Examples include swarms of cells, insects, birds, and fish, to name just a few. A common characteristic of such swarms is that the individuals consisting of swarms are passive and, therefore, cannot directly and actively control the behavior of the whole swarm~\citep{LICITRA201714374}. Furthermore, the individuals in such swarms do not necessarily admit modeling with the  second-order-integrator method~\citep[see, e.g.,][]{zhang2015leader}. Therefore, although the control of multi-agent systems with artificial objects has been studied extensively in conjunction with the development of control engineering centered on feedback control~\citep{jiang2020leader}, these methods are not necessarily effective in controlling the swarm of non-artificial agents. 

Among various objectives in swarm control, of particular  importance are \emph{guidance} and \emph{separation}; the former refers to the problem of guiding the whole swarm to a particular goal area, while the latter to the problem of splitting the swarm into multiple sub-swarms. As for the guidance of the swarms, various methodologies are available in the literature. For example, \cite{LICITRA201714374} presented a guidance methodology for a swarm of non-artificial agents using Lyapunov methods.  \cite{chipade2021multiagent} proposed a StringNet Herding algorithm to guide a swarm in an environment with dynamic obstacles. \cite{qu2021adaptive} adopted a novel reinforcement learning method for swarm guidance. \cite{chipade2019herding} presented a guidance method based on vector fields.

hsIn contrast with the aforementioned accumulating results for navigating swarms, there is a scarce of literature on the separation of swarms. Nevertheless, we can find various emerging applications in which separation control is indispensable. Such applications can be primarily found in the context of micro- and nano-particles. For example, \cite{xie2019reconfigurable} studied reconfigurable magnetic microrobot swarm, which accomplishes the micro-manipulation of different components through splitting and merging these passive microrobots.  \cite{shklarsh2012collective} provided a novel approach to carry particle cargo by the separation of a bacterial swarm, whose dynamics follows the interaction of repulsion/attraction forces. 

To fill in the gap between the aforementioned emerging applications of separation control and the research trend in the systems and control theory, in this paper, we tackle the problem of separating a specific agent from a swarm by using an external agent. We employ the so-called \emph{shepherding model}~\citep{long2020comprehensive}, which is inspired by the real shepherding behavior of sheep flock by shepherds. The shepherding model specifically refers to the situation where ``sheep'' agents avoid a ``shepherd'' agent while also interacting with each other according to the well-known Boid model~\citep[see, e.g.,][]{beaver2020beyond}. Then, our objective in this paper is to develop a movement algorithm of the shepherd to single out a specific target sheep agent from the swarm. We remark that, in contrast with the various results for navigation~\citep[see, e.g.,][]{ROSSI2018112,LICITRA201714374,go2021solving}, 
we do not find in the literature an effective separation algorithm under the shepherding framework except for the work by~\cite{goel2019leader}, where the authors studied the problem of splitting a swarm under the shepherding framework. Unfortunately, the algorithm presented by the authors does not allow us to control the size of split sub-swarms. Also, the algorithm does not distinguish the individual sheep agents in the original swarm. For these reasons, we cannot directly employ the algorithm presented in the reference for singling out a target sheep agent. 

We also investigate the connectivity maintenance of the swarm when separating the target sheep. A primary reason for this focus is that, within the applications of non-artificial swarm mentioned above, failing to maintain connectivity amongst those agents may lead to mission failure~\citep{mathews2012biologically}. In recent years, there also has been a gradual increase in the shepherding connectivity maintenance. For example, \cite{mohamed2021graph} assumes a robotic/physical implementation of shepherding where limited sensing range may deteriorate the performance of the swarm (e.g. magnetic microrobot swarm connections~\citep{kantaros2018control}.  And they proposed a connectivity-aware approach to enhance the swarm's connectivity. However, the approach to maintain the connectivity on the shepherding singling scenario remains open for investigation. 

This paper is organized as follows. In Section~\ref{sec:prob}, we formulate the problem studied in this paper. In Section~\ref{sec:two}, we present our formal analysis on the special case in which there exist only two sheep agents. Based on this analysis, we in Section~\ref{sec:main} present our algorithm for singling our a target sheep from the swarm while maintaining the connectivity in the remaining swarm. We finally present simulation results in Section~\ref{sec:sim} to illustrate the effectiveness of the proposed algorithm. 

We use the following notations in this paper. We let  $\lVert \cdot \rVert$ denote the Euclidean norm of a vector. We use the notation 
$$
    [N] = \{1, \dotsc, N\}
$$ 
for a positive integer~$N$. 

\section{Problem formulation}\label{sec:prob}

We consider the situation in which there exist $N$ agents called \emph{sheep} and one herder agent called a  \emph{shepherd} on the two-dimensional plane~$\mathbb{R}^2$. For all $i\in [N]$ and $k \geq 0$, we let $x_i(k)\in \mathbb{R}^2$ denote the position of the $i$th sheep at time $k$.
We assume the following first-order dynamics 
\begin{equation} \label{eq:x_i(k+1)=}
x_i(k+1) = x_i(k) + v_i(k). 
\end{equation}
In this equation, $v_i(k) \in \mathbb{R}^2$ denotes the movement vector of the $i$th sheep at time $k$ and is constructed as
\begin{equation} \label{eq:2}
v_i(k) = \phi_{\bar v}\left(K_{s1}v_i^1(k)+ K_{s2}v_i^2(k)+ K_{s3}v_i^3(k)\right), 
\end{equation}
where the mapping
\begin{equation*}
    \phi_{\bar v}
    \colon 
    \mathbb{R}^2\to \mathbb{R}^2
    \colon 
    x \mapsto \min(\bar v, \norm{x})\frac{x}{\norm x}
\end{equation*}
limits the maximum velocity of the agents to $\bar v > 0$,  vectors~$v_i^1(k), v_i^2(k) \in \mathbb{R}^2$ represent the movement vectors corresponding to the repulsion and attraction force in the Boids model~\citep[see, e.g.,][]{beaver2020beyond}, respectively, and $v_i^3(k)\in \mathbb{R}^2$ represents the repulsion force applied from the shepherd. The coefficients~$K_{s1}$, $K_{s2}$, and $K_{s3}$ are assumed to be nonnegative numbers. 

We assume that each sheep interacts with other agents within a finite range. We specifically assume that the $i$th sheep receives forces from the sheep
located in the  open disc
\begin{equation*}
    S_i(k) = \{z \in \mathbb{R}^2\mid \norm{z-x_i(k)} < R\}. 
\end{equation*}
The constant $R$ is called the sensing radius of the sheep. Then, we let $N_i(k)$ denote the set of indices of the sheep located in the region $S_i(k)$ at time $k$; i.e., let us define
\begin{equation*} \label{eq:3}
N_i(k) = \{j\in [N]\setminus \{i\}
\mid x_j(k)\in S_i(k)\}.
\end{equation*}
Then, the movement vectors $v_i^1(k)$, $v_i^2(k)$, $v_i^3(k)$ in \eqref{eq:2} are defined as follows:
\begin{align}
\label{4}
v_i^1(k) &= -\frac{1}{|N_i(k)|}\sum_{j\in N_i(k)}\frac{x_j(k)-x_i(k)}{\lVert x_i(k) - x_j(k)\rVert^3},    
\\
\label{5}
v_i^2(k) &= \frac{1}{|N_i(k)|}\sum_{j\in N_i(k)}\frac{x_j(k)-x_i(k)}{\lVert x_i(k) - x_j(k)\rVert},
\\
\label{6}
v_i^3(k) &= -\frac{y(k)-x_i(k)}{\lVert y(k) - x_i(k)\rVert^3}, 
\end{align}
where $y(k)\in \mathbb{R}^2$ denotes the position of the shepherd at time $k$. We remark that equations~\eqref{4}--\eqref{6} are ill-defined if the set $N_i(k)$ is empty; to avoid this potential undecidablility, we let $v_i^1(k)=v_i^2(k)=v_i^3(k) = 0$  when $N_i(k) = \emptyset$. 




In this paper, we are interested in the problem of separating a specific ``target'' sheep, labeled by $t \in [N]$, from the swarm by using the repulsion force exerted by the shepherd agent. Before precisely stating the problem, let us give a formal definition of the separation as follows. Let $k\geq 0$ be arbitrary. We say that the target sheep~$t$ is separated from the swarm at time $k$ if $N_t(k) = \emptyset$. Also, we let 
\begin{equation}\label{eq:Gtk}
    G_{t}(k) = ([N]\backslash \{t\}, E(k))
\end{equation}
denote the undirected network in which two nodes $i, j \in [N]\backslash \{t\}$ are adjacent (i.e., $\{i, j\} \in E(k)$) if and only if $\norm{x_i(k) - x_j(k)} <R$. We are now ready to state the problem studied in this paper.

\begin{prob}\label{prob:}
Let $t\in [N]$ be given.  Design the movement law of the shepherd to separate the sheep~$t$ from the swarm while keeping the network~$G_{t}(k)$ connected.
\end{prob}

\section{Two sheep-analysis}\label{sec:two}

Due to the nonlinearity of the dynamics of the sheep agents, it is not necessarily realistic to give an analytical solution to Problem~\ref{prob:} for a large swarm. Therefore, in this section, we first focus on the specific case where the swarm contains only two sheep agents. We will specifically present a sufficient condition on the dynamics of the shepherd agent for eventually separating the two sheep agents. The obtained result will be leveraged to construct our algorithm for separation in a general swarm in Section~\ref{sec:main}.

Let us assume $N=2$. For all $k\geq 0$, define the vector 
\begin{equation*}
    \Delta(k) = x_2(k)-x_1(k). 
\end{equation*}
One strategy for separating a sheep from the other is to manipulate the movement of the shepherd agent so that the distance $\norm{\Delta(k)}$ increases over time. The following proposition gives a sufficient condition for achieving the increasing monotonicity. 

\begin{prop}\label{prop:masaki}
Let $k\geq 0$ be an integer and let $c$ be a real constant. Assume that the position of the shepherd satisfies 
\begin{equation}\label{eq:y-x_1}
    y(k)-x_1(k) = c \Delta(k). 
\end{equation}
Further assume that 
\begin{equation}\label{eq:normdeltaleqR}
\norm{\Delta(k)}\leq R    
\end{equation}
holds. Then, we have 
\begin{equation*}\label{eq:Deltak<k+1}
    \norm{\Delta(k)} < \norm{\Delta(k+1)} 
\end{equation*}
if 
\begin{equation}\label{eq:cin...}
c \in {\mathcal C_1 \cup \mathcal C_2 \cup \mathcal C_3}, 
\end{equation}
where the sets $\mathcal C_1$, $\mathcal C_2$, and $\mathcal C_3 \subset \mathbb R$ are defined by 
\begin{align*}
    &\begin{multlined}[.85\linewidth]
    \mathcal C_1 = 
    \bigg\{
    c \in (-\infty, 0) \mid c^{-2}-(1-c)^{-2} >\\ \frac{2(K_{s1}-R^2K_{s2}+R^2\max(K_{s2}, R))}{K_{s3}}
    \bigg\}, 
    \end{multlined}
    \\
    &
    \begin{multlined}[.85\linewidth]
    \mathcal C_2 = 
    \bigg\{
    c \in (0, 1)\mid c^{-2}+(c-1)^{-2}>\\ \frac{2(K_{s2}R^2-K_{s1})}{K_{s3}}
    \bigg\},
    \end{multlined}
    \\
    &
    \begin{multlined}[.85\linewidth]
    \mathcal C_3 = 
    \bigg\{
    c \in (1, \infty)\mid (c-1)^{-2}-c^{-2} >\\ \frac{2(K_{s1}-R^2K_{s2}+R^2\max(K_{s2}, R))}{K_{s3}}
    \bigg\} .
    \end{multlined}
\end{align*}

\end{prop}


\begin{pf}
Define $e(k) = \norm{\Delta(k+1)}-\norm{\Delta(k)}$. We need to show 
\begin{equation}\label{eq:e(k)>0}
e(k) > 0
\end{equation}
 under the assumptions stated in the proposition. From \eqref{eq:x_i(k+1)=}, we obtain 
\begin{equation}\label{eq:e(k)=...}
    e(k) = \norm{\Delta(k) + v_2(k)-v_1(k)} - \norm{\Delta(k)}. 
\end{equation}
Because \eqref{eq:y-x_1} shows 
$y(k)-x_2(k) = (c-1)\Delta(k)$, from equations~\eqref{4}--\eqref{6} we obtain 
    $v_2(k)-v_1(k) = f(k)\Delta(k)$, 
where the scalar $f(k)$ is defined by 
\begin{equation}\label{eq:deff(k)}
\begin{multlined}[.8\linewidth]
f(k)=
\frac{cK_{s3}}{\norm{c\Delta(k)}^3}- \frac{(c-1)K_{s3}}{\norm{(c-1)\Delta(k)}^3}\\
+2\frac{K_{s1}-K_{s2}\norm{\Delta(k)}^2}{\norm{\Delta(k)}^3}. 
\end{multlined}
\end{equation}
Therefore,
equation~\eqref{eq:e(k)=...} yields
\begin{equation}\label{difference_ek}
 e(k) = \norm{\Delta(k)}(\abs{f(k)+1}-1). 
\end{equation}

Now, assume that $c$ satisfies \eqref{eq:cin...}. Then, $c$ belongs to either $\mathcal C_1$, $\mathcal C_2$, or $\mathcal C_3$. Let us first consider the first case, i.e., $c\in \mathcal C_1$. Then, the following inequalities hold:
\begin{gather*}
    2K_{s1}+((1-c)^{-2}-c^{-2})K_{s2} <0, 
    \\
    2R^3 - 2K_{s2}R^2 + 2K_{s1}+((1-c)^{-2}-c^{-2})K_{s3} <0. 
\end{gather*}
Therefore, a simple analysis shows $$
2d^3 - 2K_{s2}d^2 + 2K_{s1}+((1-c)^{-2}-c^{-2}) K_{s3} < 0
$$
for all $d\in [0, R]$. Because of the assumption~\eqref{eq:normdeltaleqR}, we obtain 
    $2\norm{\Delta(k)}^3 - 2K_{s2}\norm{\Delta(k)}^2 + 2K_{s1}+((1-c)^{-2}-c^{-2}) K_{s3} < 0$, 
which implies $f(k) < -2$ because $c<0$. Therefore, from \eqref{difference_ek}, we obtain $e(k) > 0$, as desired. 

Let us consider the second case of $c\in \mathcal C_2$. In this case, because $0<c<1$, from \eqref{eq:deff(k)} we obtain 
$$
\norm{\Delta(k)}^3f(k) = 
2K_{s1} - 2K_{s2}\norm{\Delta(k)}^2 + {K_{s3}}((1-c)^{-2} + c^{-2}),
$$ 
which is decreasing as a function of $\norm{\Delta(k)}$. By using the assumption \eqref{eq:normdeltaleqR}, we obtain 
$\norm{\Delta(k)}^3f(k) 
\geq 
2K_{s1} - 2K_{s2}R^2 + {K_{s3}}\left((1-c)^{-2} + c^{-2}\right)
\geq 
 0$
and, hence, $e(k)>0$, as desired. 

Finally, let us consider the case of $c\in \mathcal C_3$.  In this case, from \eqref{eq:deff(k)} we obtain  $\norm{\Delta(k)}^3f(k) =  2K_{s1} - 2K_{s2}\norm{\Delta(k)}^2 + {K_{s3}}(c^{-2}-(1-c)^{-2})$. Therefore, \eqref{eq:e(k)>0} and \eqref{difference_ek} yield $2\norm{\Delta(k)}^3-2K_{s2}\norm{\Delta}^2+2K_{s1}-K_{s3}(c^{-2}-(1-c)^{-2})<0$. Hence, under the assumption \eqref{eq:normdeltaleqR}, we obtain 
\begin{gather*}
    2K_{s1}-((1-c)^{-2}-c^{-2})K_{s2} <0, 
    \\
    2R^3 - 2K_{s2}R^2 + 2K_{s1}-((1-c)^{-2}-c^{-2})K_{s2} <0, 
\end{gather*}
which imply 
\begin{equation*}
\begin{multlined}
(c-1)^{-2}-c^{-2}>\frac{2(K_{s1}-R^2K_{s2}+R^2\max(K_{s2}, R))}{K_{s3}}. 
\end{multlined}
\end{equation*}
This inequality indicates $f(k)<-2$ and, therefore, we can conclude $e(k)>0$, as desired.
\end{pf}

The contribution of Proposition~\ref{prop:masaki} is in presenting a simple control law~\eqref{eq:y-x_1} for realizing an asymptotic separation of two sheep agents. Because the analysis focuses on the special case where only two sheep agents exist in the field, it is not necessarily appropriate to directly use the control law for solving Problem~\ref{prob:} for larger swarms. However, we can still expect that the control law is effective locally, both spatially and temporally. Therefore, in the next section, we leverage the control law to develop a separation algorithm applicable to swarms with general sizes. 

\section{Separation algorithm} \label{sec:main}

In this section, we propose an algorithm for solving Problem~\ref{prob:}. The algorithm is based on our analysis of the swarm of two sheep in Section~\ref{sec:two}. Within the proposed algorithm, the shepherd agent separates the target sheep from each of the other sheep agents iteratively until the target sheep separates from all the other sheep agents. 
Specifically, in this section, we first describe our algorithm for separating the target sheep, denoted by $t$,  from another sheep. We then describe our algorithm for separating the sheep~$t$ from the swarm of all the other sheep. 

\begin{algorithm}[tb]\label{alg:}
\caption{Shepherding singling}\label{algorithm}
\begin{algorithmic}[1]
\Procedure {main}{$y$,\,$x_t$,\,$x_{i}$,\,$\epsilon$,\,$\bar v$,\,$R$,\,$N_t$}
\For {sheep in $t$'s neighbors}
\State{$p\leftarrow N_t$ select $p$ from $t$'s neighbors randomly.}
\State {$y^*\leftarrow$ \Call{shepherd ideal position} {$\epsilon$,\,$x_p$,\,$x_i$}}
\State {$y\leftarrow$ \Call{path plan} {$y^*$,\,$y$,\,$\bar v$,\,$x_i$,\,$x_t$,\,$x_p$,\,$\epsilon$}}
\State{$x_p$,\,$x_v$,\,$x_i$ $\leftarrow$ dynamics($y$), update swarm; }
\While{$\norm{x_t-x_p}\leq R$}
\State {$y^*\leftarrow$ \Call{shepherd ideal position} {$\epsilon$,\,$x_p$,\,$x_i$}}
\If{shepherd can arrive $y^*$ in one step}
\State{$y\leftarrow y^*$}
\Else
\State {{$y\leftarrow$\Call{path plan} {$y^*$,\,$y$,\,$\bar v$,\,$x_i$,\,$x_t$,\,$x_p$,\,$\epsilon$}}}
\EndIf
\State{$x_p$,\,$x_v$,\,$x_i$ $\leftarrow$ dynamics($y$), update swarm; }
\EndWhile
\If{target sheep singled out from swarm}
\State \textbf{return} $y$,\,$x_t$,\,$x_i$
\EndIf
\EndFor
\EndProcedure
\Procedure{shepherd ideal position}{$\epsilon$,\,$x_p$,\,$x_i$}
\State{$y_{range}\leftarrow(x_p,x_t)$, candidate of $y$ by Proposition2}
\State{$N_p^g\leftarrow$solve \eqref{21}, extended set;}
\State {$v_p^*\leftarrow$solve \eqref{vp*}, the ideal velocity of sheep~$p$;}
\State {$y^*\leftarrow$ solve \eqref{eq:2}, shepherd's ideal position;}
  \If {$y^*$ doesn't satisfy Proposition2}
  \State {take the closest value from $Y$}
  \EndIf
\State \textbf{return} $y^*$
\EndProcedure

\Procedure{path plan}{$y$,\,$y^*$,\,$\bar v$,\,$x_i$,\,$x_t$,\,$x_p$, $\epsilon$}
\Repeat
\State{$path\leftarrow$ generate shepherd path by A* }
\State{$\norm{y(k+1)-y(k)}=\bar v$, $y(k+1) ,y(k)\in path$, update $y$ follows the path for distance of $\bar v$}
\State{$x_p$,\,$x_t$,\,$x_i$ $\leftarrow$ dynamics($y$), update swarm; }
\State{$y^*\leftarrow$\Call{shepherd ideal position} {$x_p$,\,$x_i$,\,$\epsilon$}}
the distance between\Until{$\norm{y-y^*}\leq \norm{\bar v}$}, the distance between $y$ and $y^*$ smaller than $\bar v$
\State \textbf{return} $y$
\EndProcedure
\end{algorithmic}
\vspace{1.5mm}
\end{algorithm}

\subsection{Separation from a sheep}\label{sec:separationsheep}
\label{sec:separationswarm}

{In this section, we describe our algorithm for separating the target sheep~$t$ from another sheep~$p$, which we call the ``pinning sheep''.} 
{In order to maintain and recover the fragile connections of $p$ to the other sheep ranging out of the sense radius $R$,} we first introduce the extended set of the neighbor of a sheep as
\begin{equation}\label{21}
    N_p^{g}(k) = \{j \in [N] \backslash \{p\}\mid \lVert x_p(k) - x_j(k)\rVert \leq R + \epsilon\},
\end{equation}
where $0< \epsilon <  R$ is a constant. 
We then introduce the following two scalars. The first one is concerned with the connectivity of the pinning sheep~$p$ with the swarm as well as the separation of the sheep~$p$ from the target sheep~$t$ and is defined by 
\begin{equation*}
    D_p(k) = \sum_{j\in N_p^{g}(k)}\ell_j(\lVert x_p(k) - x_j(k)\rVert-R), 
\end{equation*}
where $\ell_j \in \{+1, -1\}$ is defined by
\begin{equation*}
   \ell_j =
\begin{cases} 
+1, & \mbox{if sheep~$j$ is the target sheep,}\\
-1, & \mbox{otherwise. }
\end{cases}
\end{equation*}
Roughly speaking, 
{the function $D_p(k)$ is responsible for the maintenance of the initially formed connections between $p$ and the sheep in $N_p^g(k)$.%
}
The second quantity is concerned with the alignment of the pinning sheep~$p$ with other sheep agents and is defined by 
\begin{equation*}
    V_p(k) = \sum_{j\in N_p^{g}(k)}\ell_j \lVert v_j(k)-v_p(k)\rVert. 
\end{equation*}
Using these scalars defined above, we now define the \emph{ideal velocity} of the pinning sheep~$p$ as 
\begin{equation}\label{vp*}
\begin{split}
 v_p^{*}(k+1)= - \frac{D_p(k) + V_p(k)}{
    \sum_{j\in N_p^g(k)} \lVert x_p - x_j\rVert
    }\sum_{j\in N_p^{g}(k)}  
 (x_p(k)-x_j(k)).
\end{split} 
\end{equation}
%
Thus $v^*_p$ in \eqref{vp*} controls the pinning sheep~$p$ staying away from target sheep~$t$. And also under the control of $v^*_p$, the pinning sheep~$p$ approaches to the other sheep within $N_p^g$ except target sheep~$t$.

From \eqref{vp*}, we can find the ideal position of the shepherd as
\begin{equation}\label{y*}
\begin{split}
 &y^{*}(k+1)=\\
 &x_p(k)-\frac{v_p^*(k+1)-K_{s1}v_p^1(k+1)-K_{s2}v_p^2(k+1)}{K_{s3}\norm{y^{*}(k+1)-x_p(k)}^{-3}}.
\end{split} 
\end{equation}
This ideal position is derived from our heuristic and, therefore, moving the shepherd toward this position does not necessarily realize the desired separation. Therefore, we employ Proposition~\ref{prop:masaki} as follows. First, define the set 
\begin{equation*}
\begin{multlined}[.8\linewidth]
    Y(k) = \{z \in \mathbb{R}^2 \mid 
    z-x_p(k) = c(x_t(k)-x_p(k)),\\ c \in \mathcal C_1\cup \mathcal C_2 \cup \mathcal C_3\}.
    \end{multlined}
\end{equation*}
We then take the point in the set~$Y(k)$ that is closest to the ideal position~$y^{*}(k+1)$. The closest point is finally used as the next position of the shepherd agent. 


\subsection{Separation from swarm}

We present the whole algorithm for achieving the separation in Algorithm~\ref{algorithm}. The algorithm consists of the following three parts: i) MAIN; ii) SHEPHERD IDEAL POSITION; iii) PATH PLAN. 

MAIN starts with selecting the pinning sheep around the target sheep. Then based on this selected pinning sheep~$p$, we solve the shepherd ideal position~$y^*$ and the next step movement of the shepherd by calling the other two procedures. Then it updates all the agents' information (position and velocity) according to~\eqref{eq:x_i(k+1)=}--\eqref{6} until target sheep~$t$ lose connection to sheep~$p$.

SHEPHERD IDEAL POSITION aims to obtain the ideal position~$y^*$ according to the \eqref{21}--\eqref{y*}, and then compare the calculated result with the Proposition~\ref{prop:masaki}: if $y^*$ here doesn't satisfy the Proposition, then take the closest value from the feasible set given by the Proposition~\ref{prop:masaki} as $y^*$. 
Notice that Proposition~\ref{prop:masaki} gives the separation conditions of the shepherd position, thus the value of velocity constraint has no affects to the fact of separation. 

PATH PLAN is used to guide the shepherd moves to the ideal position~$y^*$ because of the limitation of the velocity constraint. During this process, we generate a $path$ from the current position~$y$ to $y^*$ and move the shepherd along this $path$ with its maximum velocity in each time step. All the information, including $y^*$ and the trajectory, are time-varying and updated after the movement of $y$ in each time step. Finally, the algorithm outputs the positions of the swarm once the target sheep is singled out.


\begin{figure}[tb]
\begin{center}
\includegraphics[width=6cm]{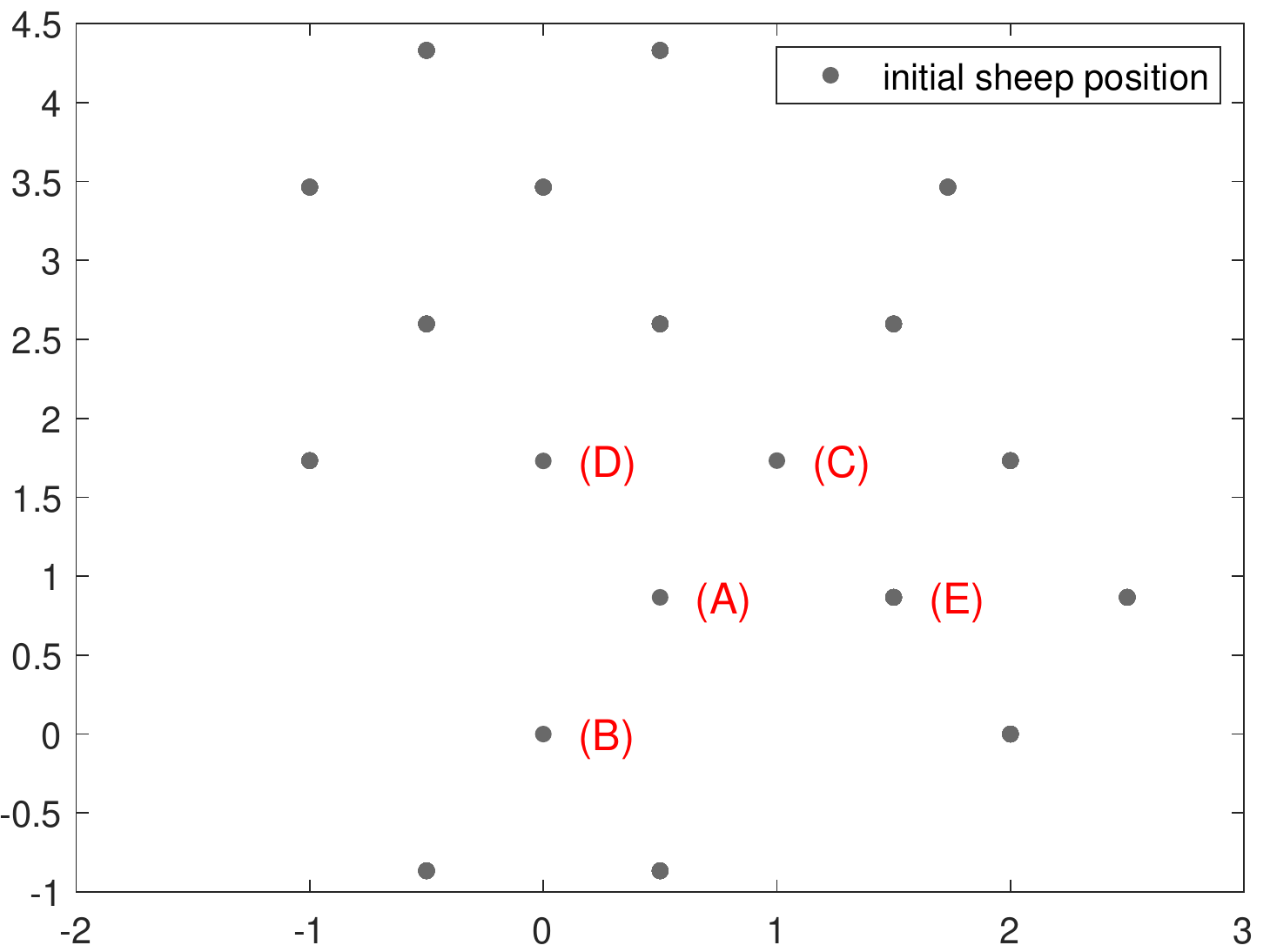}    
\caption{The target sheep in each of the experiments is $A$, $B$, $C$, $D$, and $E$, respectively}
\label{fig:inittarget}
\end{center}
\end{figure}


\renewcommand\thesubfigure{(\roman{subfigure})}

\begin{figure*}[tb]
\centering
\subfigcapskip=-2pt
\subfigure[Proposed method]{\centering
\includegraphics[width=1\linewidth]{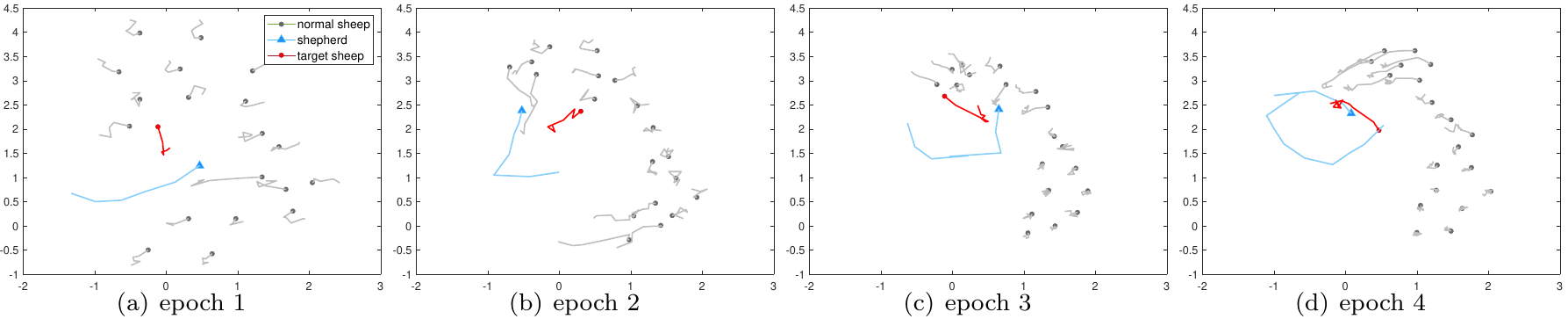}
\vspace{.75mm}
}
\\
\subfigure[Baseline method (Bipartite)]{\centering
\includegraphics[width=1\linewidth]{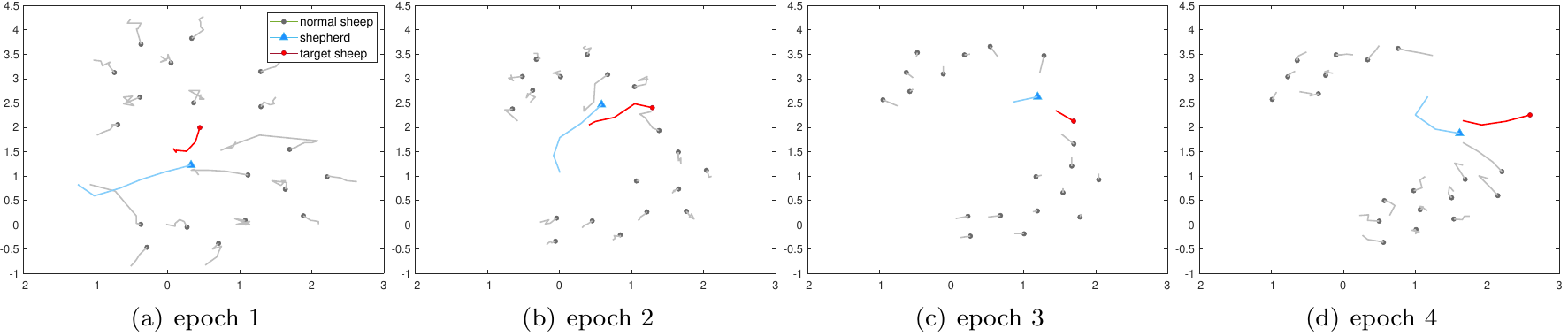}
\vspace{.75mm}
}
\vspace{-3mm}
\caption{Separation of sheep D.}
\label{normal_experiment_D}
\vspace{1mm}
\centering
\subfigcapskip=-2pt
\subfigure[Proposed method]{\centering
\includegraphics[width=1\linewidth]{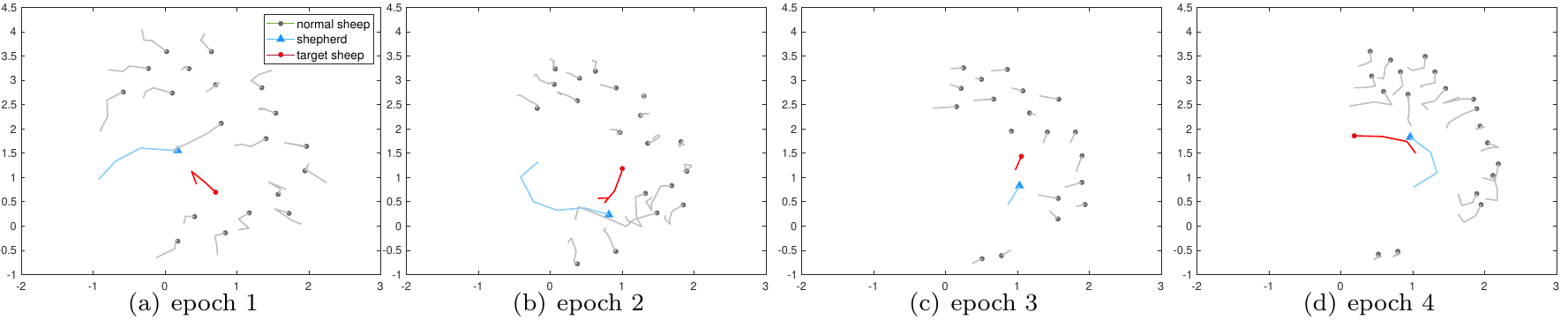}
\vspace{.75mm}
}
\\
\subfigure[Baseline method (Bipartite)]{\centering
\includegraphics[width=1\linewidth]{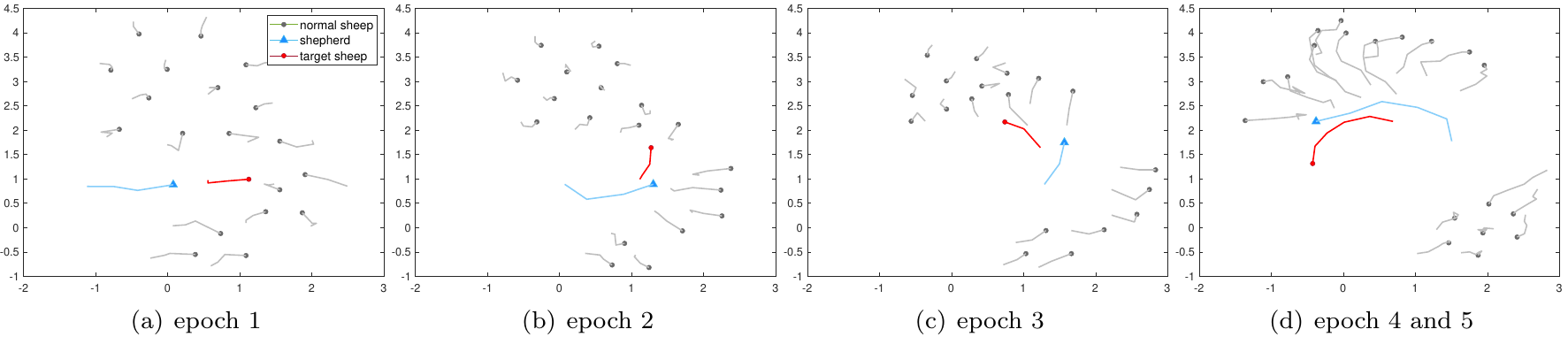}
\vspace{.75mm}
}
\vspace{-3mm}
\caption{Separation of sheep A.}
\label{normal_experiment_A}
\end{figure*}

\section{SIMULATION RESULTS and analysis}\label{sec:sim}

{This section is devoted to presenting numerical simulations of the proposed algorithm. Throughout the paper, we let $K_{s1}=1.0$, $K_{s2}=4.0$, and~$R=1.0$ to make the sheep keep an equilibrium distance~$0.5$ with each other when there is no influence from the shepherd. Considering the swarm connectivity, the shepherd should not have too large repulsion to the sheep. For this reason, we have chosen to use $K_{s3}=0.5$ and $\bar v = 0.5R$ in our experiment. As for the value of $\epsilon$, we performed preliminary experiments to find an appropriate value and have chosen $\epsilon = 0.3$.}
{We show the initial configurations of the sheep agents in  Fig.~\ref{fig:inittarget}. In our experiment, we let one of the sheep A, B, C, D, or E as the target sheep and perform its separation from the swarm.} 

{As a baseline method for comparison, we select the method presented by~\cite{goel2019leader}, where the authors use a "predator" through the middle point in a pair of sheep to do the separation. We call this method ``Bipartite" in this paper. In order to set the Bipartite method as a matched group, we replace the calculation of $y^*$ (see SHEPHERD IDEAL POSITION in Algorithm~\ref{algorithm}) as the middle point between sheep~$p$ and $t$, while keeping the other parts in the matched group experiments.
Fig.~\ref{normal_experiment_D} shows the comparison between proposed and the Bipartite method. We can observe a target sheep lose all the connections to the other sheep under the control from the shepherd. And the comparison experiment is under the same conditions and parameters where swarm splits into several subgroups.}
For further analysing the connectivity in the singling process, 
we present one typical process to single out the sheep~$A$ in Fig.~\ref{normal_experiment_A}. 

\begin{figure}[tb]
\begin{center}
\includegraphics[width=6cm]{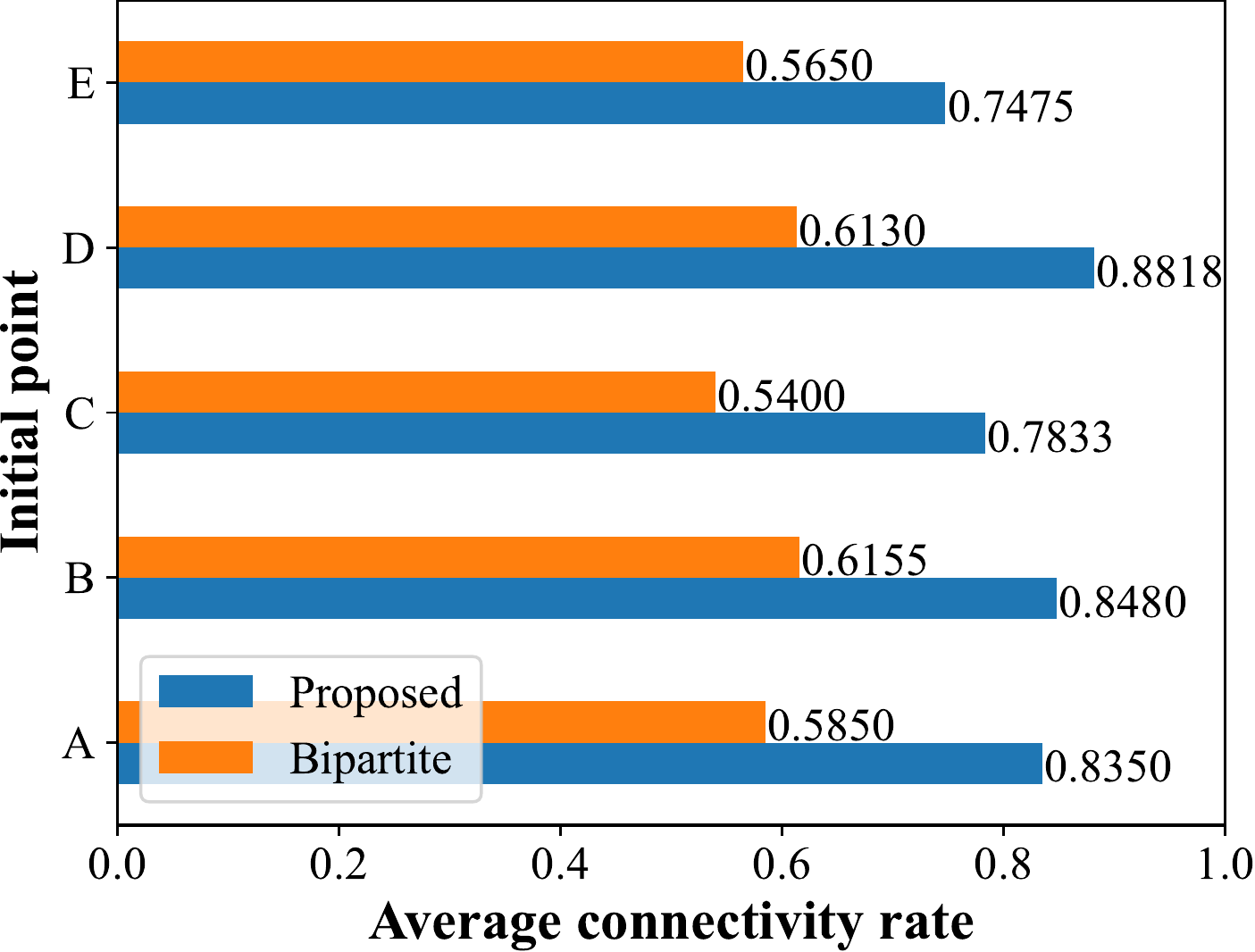}    
\caption{Average connectivity rate in different target sheep singling experiment.}
\label{cr}
\end{center}
\end{figure}

Let us quantitatively compare the proposed algorithm and the baseline method in terms of maintaining the connectivity. For this purpose, we use the size of the maximum component of the interaction network $G_t(k)$ in the remaining swarm (see \eqref{eq:Gtk} for the definition). In Fig.~\ref{cr}, we show the average size of the maximum component. Notice that, in the proposed algorithm, the pinning sheep are selected randomly. Thus we take the average of 50 experiments for each initial target sheep. We can see that our method improves the average connectivity rate, especially for singling out the target at the boundary of the swarm. This is because the feedback control to the most vulnerable connections of $p$ to the others promotes the swarm's connectivity rate to a particular extent.

On the other hand, there is a partial decrease in the connectivity rate when the target sheep moves to the deep position of the swarm. We can observe in Fig.~\ref{normal_experiment_A} that some sheep at any possible entrance have blocked the path to the ideal position~$y^*$ of the shepherd, and these sheep were influenced and lost connections. One reason is that the controllers are not rigorously proved to be convergent to the swarm's connectivity, and the velocity constraint should not be ignored.

\section{Conclusion}
In this paper, we have proposed a shepherding-based algorithm for separating a target agent from a swarm while maintaining the connectivity within the remaining swarm. In contrast to the existing non-artificial swarm separation method relying on a shepherd split down the middle, separation control and connectivity maintenance are unified in our approach, which singles out the target sheep and guarantees the other sheep's connectivity to a certain degree. This unification is  achieved partially by our analysis of the 2-sheep separation.
We have presented numerical simulations to illustrate the proposed separation algorithm. Although we have confirmed the effectiveness of the proposed algorithm, its performance is not yet investigated theoretically. Therefore, a future work includes developing a separation algorithm with theoretically guaranteed performance. 

\end{document}